\documentclass{article}

\usepackage{arxiv}

\usepackage[utf8]{inputenc} 
\usepackage[nottoc]{tocbibind}
\usepackage[T1]{fontenc}    
\usepackage{hyperref}       
\usepackage{url}            
\usepackage{booktabs}       
\usepackage{amsfonts}       
\usepackage{nicefrac}       
\usepackage{microtype}      
\usepackage{cite}
\usepackage{amsmath,amssymb,amsfonts}
\usepackage{algorithmic}
\usepackage{graphicx}
\usepackage{textcomp}
\usepackage{xcolor}
\usepackage{hyperref}
\usepackage{tcolorbox}
\usepackage{float}

\title{Continuous Systematic Literature Review: An approach for open science}

\author{ 
  Bianca M. Napoleão\\
  Laboratoire d’informatique formelle \\
  Université du Québec à Chicoutimi\\
  \texttt{bianca.minetto-napoleao1@uqac.ca} \\
   \And
  Fabio Petrillo\\
  Laboratoire d’informatique formelle \\
  Université du Québec à Chicoutimi\\
  \texttt{fabio@petrillo.com} \\
    \And
  Sylvain Hallé\\
  Laboratoire d’informatique formelle \\
  Université du Québec à Chicoutimi\\
  \texttt{shalle@acm.org} \\
}

\hypersetup{
pdftitle={},
pdfsubject={q-bio.NC, q-bio.QM},
pdfauthor={David S.~Hippocampus, Elias D.~Striatum},
pdfkeywords={First keyword, Second keyword, More},
}

\begin{document}
\maketitle

 \begin{abstract}
	Systematic Literature Reviews (SLRs) play an important role in the Evidence-Based Software Engineering scenario. With the advance of the computer science field and the growth of research publications, new evidence continuously arises.  This fact impacts directly on the purpose of keeping SLRs up-to-date which could lead researchers to obsolete conclusions or decisions about a research problem or investigation.  Creating and maintaining SLRs up-to-date demand a significant effort due to several reasons such as the rapid increase in the amount of evidence, limitation of available databases and lack of detailed protocol documentation and data availability.
	Conventionally, in software engineering SLRs are not updated or updated intermittently leaving gaps between updates during which time the SLR may be missing important new research. In order to address these issues, we propose the concept, process and tooling support of Continuous Systematic Literature Review (CSLR) in SE aiming to keep SLRs constantly updated with the promotion of open science practices. This positional paper summarizes our proposal and approach under development.  

\end{abstract}


\section{Introduction}
\label{sec:introduction}

Systematic Literature Reviews (SLRs) are considered the pillar of evidence-based medicine \cite{Munn2018}. An SLR aims to identify, select, summarize and synthesize existing evidence about a research topic or phenomenon of interest \cite{Kitchenham15}.
Since the introduction of SLR in the Software Engineering (SE) field in 2004 \cite{Kitchenham04}, especially over the last years, the number of SLR has been increased substantially \cite{Kitchenham15, Silva10}.

In the context of SE, an SLR is classified as a secondary study which includes also Systematic Mappings (SMs). An SM is a kind of lightweight SLR that aims to survey the available knowledge about a topic \cite{Kitchenham15}. They follow the same process of conduction and they are considered key to the Evidence-Based Software Engineering (EBSE) field \cite{Kitchenham15}. For these reasons, both types of secondary studies will be addressed in our research. In order to facilitate the understating, in this study the SLR name will make reference to SLRs and SMs.

One known challenge in evidence-based disciplines is to keep SLR updated. 
As stated in the Cochrane's handbook \cite{Cochraine2019}, SLR that are not maintained may become out of date or misleading. With the advance of the computer science field and the growth of research publications, new evidence continuously arises. This fact impacts directly on the purpose of keeping SLRs up-to-date, consequently could leading researchers to obsolete conclusions or decisions about a research topic \cite{Watanabe20}. 

In the medicine field, SLR update has a consolidated process \cite{MedMoher2008, MedShekelle2011}. Also in SE, there are several initiatives on SLR updates such as on establishing the SLR update process \cite{Dieste08a, Mendes2020}, searching for new/updated evidence \cite{felizardo16, Wohlin2020}, selecting updated evidence \cite{Watanabe20} and experience reports \cite{Garces17, Felizardo20}. Despite the effort of the SE community to keep SLR updated, a recent systematic mapping \cite{Nepomuceno2019} showed that only 22 SLRs in SE were updated.  

Creating and maintaining SLRs up-to-date demand a significant effort for reasons such as the rapid increase in the amount of evidence \cite{Zhang18, Stol15} and the limitation of available databases \cite{Imtiaz13}. Furthermore, the lack of detailed protocol documentation and data availability make the SLR update process even more difficult since most of the tacit knowledge from the SLR conduction is lost \cite{Felizardo2020}. 

Conventionally, in SE SLRs are not updated or updated intermittently \cite{Wohlin2020}. Intermittent updating leaves gaps between updates during which time the SLR may be missing important new research, placing it at risk of inaccuracy and wasting the potential contribution of new research to evidence synthesis and decision-making.
In the medicine field, in order to mitigate the SLR updating issue, Elliott \textit{et al. }\cite{Elliott17} introduced the concept of ``Living Systematic Review'' (LSR). An LSR is an SLR that is that is continually updated, incorporating relevant new evidence as it becomes available.


In the context of software development, DevOps concept (mindset, practices and tools) brought several benefits such as: faster release of features, improvement on the monitoring of systems in production, stimulate collaboration among team members, among others. All of these benefits lead to to high quality software \cite{Bass2015}. Continuous Integration (CI), a DevOps practice, aims to build and integrate all working versions of the software code together, keeping the software updated. It includes automation, for example, a build service; and a cultural mindset such as integrate changes constantly \cite{Humble2010}.

Over the last years, the SE community started the open science movement which consists of making any research artifact available to the public addressing open access, open data and open source practices \cite{Mendez2020}. Open science approaches directly impact the SLR' conduction not limited to the access and availability of primary evidence for the conduction of SLRs, but on the adoption of open science practices during the conduction of secondary studies that reflects on the reproducibility and maintainability of the SLRs. Concepts of open science must be addressed by researchers that conduct any kind of evidence-based study.






Inspired by LSR from evidence-based medicine and considering the DevOps movement especially the CI process and practices as well as Science 2.0 practices (open science, cooperation and incremental science) \textbf{we propose the concept, process and tooling support on Continuous Systematic Literature Review (CSLR) in SE aiming to keep secondary studies constantly updated.}

This \textbf{positional paper} presents our approach named Continuous Systematic Review. This approach may collaborate with the Evidence-Based Software Engineering area, enabling an innovative way to conduct secondary studies and supporting trustworthy and up-to-date evidence for secondary studies in SE.

\section{Proposal}
\label{sec:proposal}

The update process does not have to be considered just after the SLR publication, but during the conduction of the SLR. Our research goal is to \textbf{propose and validate the process of CSLR, including guidelines and automated tooling support.} 

Our research addresses the three hypothesis detailed next. 

\begin{tcolorbox}[float, colframe=gray!25, coltitle=black, arc=0mm, title=\textbf{Hypothesis 1}]
CSLR mitigates the intermittent updating problem (secondary studies’ publication timing) that reflects on the missing of important new research. 
\end{tcolorbox}

\begin{tcolorbox}[float, colframe=gray!25, coltitle=black, arc=0mm, title=\textbf{Hypothesis 2}]
CSLR contributes to reduce the risk of inaccuracy and wasting of potential contribution of new research to evidence synthesis and decision-making.
\end{tcolorbox}

\begin{tcolorbox}[float, colframe=gray!25, coltitle=black, arc=0mm, title=\textbf{Hypothesis 3}] 
CSLR promotes open science in practice in the context of evidence-based software engineering. 
\end{tcolorbox}

In order to reach our research goals and validate our hypothesis, our research project is divided into five major steps: 1- Since the core for our approach address the search and selection of evidence activities, the first step was to performed a systematic investigation on automated support for these SLR activities \cite{Napoleao2020a}; 2- Creation and development of the CSLR process; 3- Elaboration and execution of a proof of concept validation of the CSLR process and practices; 4- Development of the automated tooling support prototypes; and 5- A complete validation of the CSLR process including the tooling support.

\section{Conclusion}

In this paper, we present our approach of Continuous Systematic Review in SE. In addition, we describe our proposal and the major steps of our research project. 

We believe that creating and introducing CSLR process, practices and tooling will contribute to mitigate the intermittent updating problem reflecting on the reduction of the risk of inaccuracy and wasting of potential contribution of new research to evidence synthesis and decision-making and will promote Science 2.0 approaches such as open science in practice on the context of EBSE.


\bibliographystyle{unsrt}
\bibliography{references}

\begin{thebibliography}{10}

\bibitem{Munn2018}
Zachary Munn, Cindy Stern, Edoardo Aromataris, Craig Lockwood, and Zoe Jordan.
\newblock What kind of systematic review should i conduct? a proposed typology
  and guidance for systematic reviewers in the medical and health sciences.
\newblock {\em BMC Medical Research Methodology}, 18, 12 2018.

\bibitem{Kitchenham15}
B.A. Kitchenham, D.~Budgen, and P.~Brereton.
\newblock {\em Evidence-Based Software Engineering and Systematic Reviews}.
\newblock Chapman \& Hall/CRC Innovations in Software Engineering and Software
  Development Series. Chapman \& Hall/CRC, 2015.

\bibitem{Kitchenham04}
B.~Kitchenham.
\newblock Procedures for performing systematic reviews.
\newblock Joint Technical Report TR/SE-0401 (Keele) - 0400011T.1 (NICTA),
  Software Engineering Group - Department of Computer Science - Keele
  University and Empirical SE - National ICT Australia Ltd, 2004.

\bibitem{Silva10}
F.Q.B. da~Silva, A.L.M. Santos, S.~Soares, A.C.C. Fran\c{c}a, and C.V.F.
  Monteiro.
\newblock Six years of systematic literature reviews in software engineering:
  an extended tertiary study.
\newblock In {\em ICSE}, pages 1--10, " ", 2010. IEEE Computer Society.

\bibitem{Cochraine2019}
Julian Higgins, James Thomas, Jackie Chandler, Miranda Cumpston, Tianjing Li,
  Matthew Page, and Vivian Welch.
\newblock Cochrane handbook for systematic reviews of interventions.
\newblock {\em Cochrane Handbook for Systematic Reviews of Interventions}, 09
  2019.

\bibitem{Watanabe20}
Willian~Massami Watanabe, Katia~Romero Felizardo, Arnaldo Candido,
  Erica~Ferreira de~Souza, Jo\ ao~Ede~de Campos~Neto, and
  Nandamudi~Lankalapalli Vijaykumar.
\newblock Reducing efforts of software engineering systematic literature
  reviews updates using text classification.
\newblock {\em Information and Software Technology}, 128:106395, 2020.

\bibitem{MedMoher2008}
David Moher, Alexander Tsertsvadze, Andrea Tricco, Martin Eccles, Jeremy
  Grimshaw, Margaret Sampson, and Nick Barrowman.
\newblock When and how to update systematic reviews.
\newblock {\em Cochrane database of systematic reviews (Online)}, 1:MR000023,
  02 2008.

\bibitem{MedShekelle2011}
Shekelle PG, Sydne Newberry, Helen Wu, Marika Booth, Aneesa Motala, Yee~Wei
  Lim, Ethan Balk, Mei Chung, Yu~WW, Lee J, Gaylor JM, Moher D, Mohammed
  Ansari, Skidmore R, and Chantelle Garritty.
\newblock {\em Identifying signals for updating systematic reviews: a
  comparison of two methods}.
\newblock 06 2011.

\bibitem{Dieste08a}
O.~Dieste, M.~L\'opez, and F.~Ramos.
\newblock Formalizing a systematic review updating process.
\newblock In {\em 6$^{th}$ Int. Conference on Software Engineering Research,
  Management and Applications (SERA'08)}, pages 143--150, 2008.

\bibitem{Mendes2020}
E.~Mendes, C.~Wohlin, K.R. Felizardo, and M.~Kalinowski.
\newblock When to update systematic literature reviews in software engineering.
\newblock {\em Journal of Systems and Software}, 167:110--167, 2020.

\bibitem{felizardo16}
Katia Felizardo, Emilia Mendes, Marcos Kalinowski, Érica~Ferreira Souza, and
  Nandamudi Vijaykumar.
\newblock Using forward snowballing to update systematic reviews in software
  engineering.
\newblock In {\em ESEM}, 2016.

\bibitem{Wohlin2020}
Claes Wohlin, Emilia Mendes, Katia Felizardo, and Marcos Kalinowski.
\newblock Guidelines for the search strategy to update systematic literature
  reviews in software engineering.
\newblock {\em Information and Software Technology}, 09 2020.

\bibitem{Garces17}
Lina Garcés, Katia Felizardo, Lucas Oliveira, and Elisa Nakagawa.
\newblock An experience report on update of systematic literature reviews.
\newblock pages 91--96, 07 2017.

\bibitem{Felizardo20}
Katia~R. Felizardo, Érica~F. de~Souza, Tamiris Malacrida, Bianca~M. Napoleão,
  Fabio Petrillo, Sylvain Hallé, Nandamudi~L. Vijaykumar, and Elisa~Y.
  Nakagawa.
\newblock Knowledge management for promoting update of systematic literature
  reviews: An experience report.
\newblock In {\em 2020 46th Euromicro Conference on Software Engineering and
  Advanced Applications (SEAA)}, pages 471--478, 2020.

\bibitem{Nepomuceno2019}
Vilmar Nepomuceno and Sergio Soares.
\newblock On the need to update systematic literature reviews.
\newblock {\em Information and Software Technology}, 109:40--42, 2019.

\bibitem{Zhang18}
L.~Zhang, Jia-Hao Tian, Jing Jiang, Y.~Liu, Meng-Yuan Pu, and Tao Yue.
\newblock Empirical research in software engineering — a literature survey.
\newblock {\em J. of Computer Science and Technology}, 33:876--899, 2018.

\bibitem{Stol15}
Klaas-Jan Stol and Brian Fitzgerald.
\newblock A holistic overview of software engineering research strategies.
\newblock In {\em CESI}, page 47–54. IEEE Press, 2015.

\bibitem{Imtiaz13}
S.~Imtiaz, M.~Bano, N.~Ikram, and M.~Niazi.
\newblock A tertiary study: Experiences of conducting systematic literature
  reviews in software engineering.
\newblock In {\em EASE}, pages 177--182. ACM, 2013.

\bibitem{Felizardo2020}
Katia Felizardo, Erica Souza, Tamiris Malacrida, Bianca Napoleão, Fabio
  Petrillo, Sylvain Hallé, Nandamudi Vijaykumar, and Elisa Nakagawa.
\newblock Knowledge management for promoting update of systematic literature
  reviews: An experience report.
\newblock pages 471--478, 08 2020.

\bibitem{Elliott17}
Julian Elliott, Anneliese Synnot, Tari Turner, Mark Simmonds, Elie Akl, Steve
  Mcdonald, Georgia Salanti, Joerg Meerpohl, Harriet MacLehose, John Hilton,
  Ian Shemilt, James Thomas, Thomas Agoritsas, Rebecca Hodder, and Juan
  Yepes-NuÃ±ez.
\newblock Living systematic review 1: Introduction - the why, what, when and
  how.
\newblock {\em Journal of Clinical Epidemiology}, 91, 09 2017.

\bibitem{Bass2015}
Len Bass, Ingo Weber, and Liming Zhu.
\newblock {\em DevOps: A Software Architect's Perspective}.
\newblock 05 2015.

\bibitem{Humble2010}
Jez Humble and David Farley.
\newblock {\em Continuous Delivery: Reliable Software Releases through Build,
  Test, and Deployment Automation}.
\newblock Addison-Wesley Professional, 1st edition, 2010.

\bibitem{Mendez2020}
Daniel Mendez, Daniel Graziotin, Stefan Wagner, and Heidi Seibold.
\newblock {\em Open Science in Software Engineering}, pages 477--501.
\newblock Springer International Publishing, Cham, 2020.

\bibitem{Napoleao2020a}
Bianca~M. Napoleão, Fabio Petrillo, and Sylvain Hallé.
\newblock Automated support for searching and selecting evidence in software
  engineering: A cross-domain systematic mapping.
\newblock In {\em Manuscript accepted in the proceedings of the 47th Euromicro
  Conference on Software Engineering and Advanced Applications (SEAA)}, 2021.

\end{thebibliography}

\end{document}